\documentclass[aps,pra,twocolumn,showpacs,notitlepage]{revtex4-1}

\usepackage{graphicx}

\begin{document}

\title{On the origin of randomness in quantum mechanics}

\author{Holger F. Hofmann}
\email{hofmann@hiroshima-u.ac.jp}
\affiliation{
Graduate School of Advanced Sciences of Matter, Hiroshima University,
Kagamiyama 1-3-1, Higashi Hiroshima 739-8530, Japan}

\begin{abstract}
Quantum statistics originate from the physics of state preparation. It is therefore wrong to think of quantum states as fundamental. In fact, quantum states are merely summaries of dynamical processes that randomize the properties of the system by drawing on the inexhaustible reservoir of quantum fluctuations provided by the physical tools used to control the quantum system. The mathematical form of the ``state vector'' is actually an expression of the laws of causality which describe the relations between physical properties in terms of the action of transformations. These laws of causality directly associate the macroscopic effects of a physical property in an interaction with the environment with dynamical changes to the system caused by the microscopic properties of that environment. 
\end{abstract}


\maketitle

\section{Introduction}

For all practical purposes, quantum mechanics is a theory of probabilities. This may appear to be unproblematic since probabilities and statistics can be studied in the context of familiar random processes such as Brownian motion or dice games. However, quantum mechanics should include classical physics as an approximation. It is therefore necessary to explain the origin of quantum fluctuations in a way that is consistent with classical determinism. 

The recent discussion of quantum information has resulted in a renewed interest in ``classical'' models of statistics, e.g. Bohmian trajectories \cite{Wis07,Hal14} and positive Wigner functions \cite{Bar12}. While these models highlight the similarities between quantum statistics and classical statistics, they fail to explain where the initial probability distributions come from. Since quantum dynamics are fully deterministic, providing a reversible map between initial properties and final properties for all possible scenarios, it should be possible to identify the origin of randomness in the physical processes associated with quantum state preparation.

Unfortunately, textbook quantum mechanics tends to gloss over the state preparation process, usually expecting the reader to accept mathematical symbols of unexplained meaning as descriptions of physical situations. It might be time to wake up to the fact that this is bad science: a physical situation is known from experience, and a proper description of a physical situation must relate to that experience. To understand physics in the real world, we need to remember that the Hilbert space vector has no physical meaning in its own right. The physics described by Hilbert space vectors should be explained as an effect of the actual physical processes that resulted in the situation they describe. 

In the following, I will show that quantum mechanics provides us with well-defined laws of causality that can be used to trace the origin of measurement results back to the process by which the quantum state was prepared. When expressed in terms of observable physical properties, these laws of causality appear in the form of complex conditional probabilities, where the complex phases are expressions of the action of transformations between the physical properties \cite{Hof11,Hof12,Hof14}. It can be shown that any interaction that selects a physical property $a$ from an initial situation with a well-defined value of $b$ dynamically randomizes $b$ according to the deterministic laws of motion described by the complex conditional probabilities. Ultimately, the randomness of this dynamical process can be traced back to microscopic properties of the environment that are beyond our control. 

In many discussions of quantum mechanics, it is assumed that the world of our experience is ``classical''. However, this is only as valid as the claim that the world of our experience is non-relativistic. The truth is that we are not aware of the huge amount of quantum fluctuations in the world around us because we are not sufficiently sensitive to them. Quantum physics is fully deterministic, but the laws of causality limit our ability to control the world around us. Whenever we control a process in the real world, we only control a small part of the physical systems involved. Quantum physics, which governs the relations between all physical properties, explains that for every macroscopic property we control, there must be a microscopic property that has no observable effects and left no trace of itself in the environment. Whenever we use a macroscopic device to impose control on the quantum world, these microscopic properties will randomize the dynamics of the quantum system along a trajectory defined by the physical property we have brought under control, resulting in the familiar predictions of randomness associated with quantum statistics. 

\section{State preparation as a process}

Quantum state preparation takes many forms and depends strongly on the technological possibilities in a given field of research. That is precisely why it has received so little attention from theorists: it seems to be a very technical problem, with unclear general implications. However, there is one procedure of quantum state preparation that is comparatively well studied, and that is the preparation of a quantum state by a projective measurement. In this process, the measurement determines a property $a$ of the system, thereby ensuring that future measurements of the property $a^\prime$ result in $a^\prime=a$ with close to 100 \% certainty. At the same time, quantum theory demands that the probabilities of a different property $b$ will be given by $P(b|a) = |\langle b \mid a \rangle|^2$, no matter what the details of the measurement process were. 

In the preparation by measurement, the randomness of the property $b'$ is a result of the disturbance of $b$ in the measurement of $a$. This randomness can be traced back to the measurement system using the standard model of a bilinear interaction. If $\hat{A}$ is the operator associated with the property $a$, the measurement interaction can be modeled by the Hamiltonian
\begin{equation}
\hat{H} = g \hat{A} \hat{p},
\end{equation}
where $\hat{p}$ is the physical property of the meter system that generates the observable shift in the meter observable. Since $\hat{p}$ generates the dynamics of the meter observable $\hat{x}$, the two do not commute. Moreover, a successful read-out of the measurement result is only possible if the initial uncertainty of $\hat{x}$ is much smaller than the differences between the meter shifts induced by different eigenvalues of $\hat{A}$. The result is that a measurement of specific outcomes $a$ requires a minimal uncertainty of $\hat{p}$, so that the dynamics of the system in the measurement interaction occurs at a random rate.

Much confusion is caused by the fact that the standard quantum model of the measurement interaction results in an entangled state. However, it is possible to first focus on the causality relation that links the randomness of the system dynamics to the initial property $\hat{p}$ of the meter system. If we treat $\hat{p}$ as a classical random parameter, then the unitary acting on the system can be written as
\begin{equation}
\hat{U} = \exp(-i \phi \hat{A}),
\end{equation}
where $\phi = g p t/\hbar$ is a phase parameter that re-scales the interaction time based on the random value of $p$. In other words, the randomness of $\hat{p}$ in the environment gives rise to a randomness in the dynamics of the system. 

In principle, the randomization of the dynamics generated by $\hat{A}$ is well known from measurement theory. However, it is usually described as ``dephasing'' or as ``decoherence,'' which hides the actual physics of the randomized unitary dynamics. Specifically, the dephasing process in a projective measurement should be understood as a randomization of the transformation parameter $\phi$. This is important because we can in principle trace back any final result of $b$ obtained after the measurement interaction to a specific combination of the meter observable $p$ with physical properties of the system before the measurement interaction. 

\section{Quantum ergodicity}

The description of causality by unitary operators is not satisfactory, since these transformations do not describe the evolution of physical properties. A proper description of causality should express the relations between physical properties, not just the changes to their statistics. As I showed recently \cite{Hof12}, such a description of causality is given by the complex probability representation of quantum states defined by the expectation values of projector products,
\begin{equation}
\label{eq:jprob}
\rho(a,b) = \langle b \mid a \rangle \langle a \mid \hat{\rho} \mid b \rangle.
\end{equation}
The advantage of this representation is that it is rather flexible with regard to the choice of basis states. It is therefore possible to interpret unitary transformations as a simple change of representation. Using the notation $\hat{U}^\dagger \mid b \rangle = \mid U(b) \rangle$,
\begin{equation}
\label{eq:dynamics}
\rho(a,U(b)) = \sum_{b^\prime} P(U(b) | a, b^\prime ) \rho(a,b^\prime), 
\end{equation}
where
\begin{eqnarray}
\label{eq:causality}
\lefteqn{P(U(b) | a, b^\prime ) = \frac{\langle b^\prime \mid \hat{U}^\dagger \mid b \rangle \langle b \mid \hat{U} \mid a \rangle }{\langle b^\prime \mid a \rangle}}
\nonumber 
\\
&=& \frac{\langle b \mid a \rangle}{\langle b^\prime \mid a \rangle} 
\sum_{a^\prime} \langle b^\prime \mid a^\prime \rangle \langle a^\prime \mid b \rangle 
e^{(-i \phi (A_a - A_{a^\prime}))}.
\end{eqnarray}
Here, the eigenvalues of $\hat{A}$ for an outcome $a$ are represented by $A_a$, since it is often useful to distinguish between qualitative results $a$ and the physical quantities represented by the eigenvalues of the operator. 

We can now introduce a quantum mechanical definition of ergodic randomization. As discussed above, the dynamics induced by a precise measurement of $\hat{A}$ corresponds to a randomization of the phase $\phi$ in the time evolution of the system. Since Eq.(\ref{eq:dynamics}) is linear in $P(U(b)| a, b^\prime )$, we can apply this average directly to the conditional probabilities, resulting in a simple description of ergodic randomization for complex joint probabilities,
\begin{equation}
\label{eq:ergodic}
P_{\mathrm{ergodic}}( U(b) | a, b^\prime ) = |\langle b \mid a \rangle|^2. 
\end{equation}
Specifically, ergodic averaging eliminates all contributions with $a \neq a^\prime$, which is mathematically equivalent to the elimination of off-diagonal elements in the density matrix. 

Eq.(\ref{eq:ergodic}) shows that the probability distribution $P(b|a)=|\langle b \mid a \rangle|^2$ is indeed the result of an ergodic randomization of the dynamics generated by $\hat{A}$ \cite{Hof14}. Specifically, the application to any initial state $\rho(a,b)$ results in 
\begin{eqnarray}
\label{eq:dynamics}
\rho(a,U(b)) &=& P(b|a) \sum_{b^\prime} \rho(a,b^\prime) 
\nonumber \\
&=& P(b|a) \langle a \mid \hat{\rho} \mid a \rangle.
\end{eqnarray}
State preparation can be explained as a sequence of ergodic randomization followed by a ``classical'' selection of the result $a$. Importantly, the physics of quantum states is the physics of ergodic ensembles. It is therefore possible to explain the origin of randomness in quantum states in terms of actual physics, without any reference to hypothetical realities within the system. 

\section{The physics of control}

It may be useful to illustrate the implications of the analysis of quantum ergodic averaging by applying it to a selection of particularly familiar examples. Perhaps the most simple example is that of a single slit. Specifically, it is possible to prepare an eigenstate of position $\mid x_q \rangle$ by passing a quantum particle through a slit at that position. As a result, the transverse momentum $\hat{p}_q$ will be randomized. Importantly, this randomization is not the result of some mysterious ``collapse'' of the wavefunction. Instead, a particle passing through the slit experiences the effects of a potential in space, resulting in a transfer of momentum from the slit to the particle. The Hamiltonian of the interaction depends on the difference between the position of the particle and the position of the slit, $x_q-x_s$. Therefore the interaction conserves the total momentum, and the momentum transfer could be monitored by measuring the momentum of the slit before and after the interaction. However, the slit is a macroscopic object of considerable mass, and the slit position can only be controlled with sufficient precision because this solid object can be fixed in place by mechanical means. Importantly, this does not mean that the slit is a ``classical'' object. The interaction with the particle is fully quantum mechanical and the source of the momentum uncertainty of the particle are the quantum fluctuations of the momentum of the screen. It is therefore essential that the measurement apparatus is not ``classical'' either. Indeed, there is no such thing as a ``classical'' object, and the more macroscopic an object gets, the more microscopic quantum fluctuations it will contribute to its interactions. 

A similar situation exists in the preparation of the polarization state of a single photon using a polarization beam splitter. In that case, the interaction with the beam splitter is conditioned by the polarization, resulting in the transfer of angular momentum from the beam splitter to the photon. The precise linear polarization selected by the beam splitter depends on the angular orientation of the beam splitter, which means that there must be a corresponding amount of uncertainty in its angular momentum, and the need to fic the axes of the beam splitter in place makes it impossible to sufficiently control the angular momentum of the beam splitter before and after the photon passes it. 

Slits and beam splitters are macroscopic devices, and a detailed quantum mechanical description of all the vibrational and rotational degrees of freedom would be difficult to formulate. However, the universality of quantum mechanics requires that all interaction processes follow the same principles. It is therefore possible to gain fundamental insights into the interaction between macroscopic and microscopic objects in the limit where the macroscopic meter system is still small enough to be represented by a single degree of freedom. This is the case in the Stern-Gerlach experiment, where the motion of an atom in an inhomogeneous magnetic field serves as the meter for a measurement of a spin component. Since this is a well-studied scenario, I will not go into detail here \cite{BasDal}. Importantly, the discussion above applies here as well: the quantum fluctuations of the meter result in a random angle of the spin precession caused by the inhomogeneous magnetic field. It is therefore possible to identify the physical origin of the randomness of subsequent spin measurements for components orthogonal to the one selected in the preparation. As in all other cases, the randomness of the quantum state is a dynamical average along a trajectory - a quantum ergodic average, and not a ``distribution of realities''.

Finally, it may be good to recall that a number of state preparation methods are actually based on cooling. It is here that the historical relation between the quantum ergodicity of state preparation and Boltzmann's original use of the concept in thermodynamics \cite{Bol1898}. As pointed out by Boltzmann, thermalization results in ensemble averages that correspond to the time-averaged motion of the system. Interestingly, it is not entirely clear what conditions are sufficient to ensure that an interaction between two systems results in a thermalization. It may therefore be worthwhile to study the cooling processes used in quantum mechanics in more detail. In all cases, one should expect that the eventual quantum fluctuations of a thermal ground state are the result of well-defined interactions with corresponding quantum fluctuations in the environment.

\section{Time-symmetry and laws of causality}

Events in the past determine the physical properties of a system in the future, and events in the future determine the physical properties of a system in the past. Physics provides the universal set of rules by which the physical properties are related to each other in the course of time, and these universal laws of physics allow us to make sense of the observable sequence of events. The ``state'' of a system is merely a specific situation among many and needs to be understood as a representation of the more general laws that are necessarily ``state-independent''. 

In quantum mechanics, the difficulty of identifying the universal laws of physics arises because state preparation and measurement are always incomplete, since it is impossible to isolate the causality of the system from the effects of the environment. However, there is a way out: weak measurements are sensitive to the physical properties of a system during its free evolution between the observable effects associated with preparation and measurement \cite{Aha88}. This is why weak measurements can be used as an empirical foundation for a time-symmetric formulation of causality. As pointed out in \cite{Hof12,Hof14}, the complex-valued conditional probabilities $P(m|a,b)$ that are defined by the weak values of the projector $\mid m \rangle \langle m \mid$ for an initial state $\mid a \rangle$ and a final state $\mid b \rangle$ express the deterministic law of causality that relates the physical property $m$ to the set of properties $(a,b)$. As explained in \cite{Hof12}, this relation replaces the classical relations $m=f_m(a,b)$, where the value of $m$ is a function of the values of $a$ and $b$. These classical relations are never fundamental: they emerge only as an approximation obtained by coarse graining over intervals corresponding to actions greater than $\hbar$. 

For a measurement of the physical property $m$, the outcome is fully determined by $(a,b)$ according to 
\begin{equation}
P_{\mathrm{exp.}}(m) = \sum_{a,b} P(m|a,b) \rho(a,b).
\end{equation}
Here, the initial state is expressed by the complex-valued joined probability $\rho(a,b)$ given in Eq.(\ref{eq:jprob}). Importantly, quantum ergodicity states that $(a,b)$ does not represent an elementary reality. Instead, $a$ and $b$ are related to each other by quantum ergodic dynamics, as expressed in the complex phases of the deterministic probability relation $P(m|a,b)$, which represent actions of transformation \cite{Hof11}. 

Time-reversibility can now be illustrated by expressing $\rho(a,b)$ as an ergodic average of $m^\prime$. The joint probability can be written in the $(m,b)$ representation as
\begin{equation}
\rho(m,b|m^\prime) = P(b|m) \delta_{m,m^\prime}.i
\end{equation}
The $(a,b)$ representation can then be obtained by using the deterministic relation between $a$ and the pairs $(m,b)$ to replace $m$ with $a$ \cite{Hof12},
\begin{eqnarray}
\rho(a,b|m^\prime) &=& \sum_m  P(a|m,b) \rho(m,b|m^\prime)
\nonumber \\
&=& P(a|m^\prime,b) P(b|m^\prime).
\end{eqnarray}
We therefore find that the causality relation that connects a preparation of $m$ to a measurement of $m$ is given by
\begin{equation}
P_{\mathrm{exp.}}(m|m^\prime) = \sum_{a,b} P(m|a,b) P(a|m^\prime,b) P(b|m^\prime).
\end{equation}
This relation is actually independent of the choice of $b$ and expresses the time reversal symmetry of state preparation and measurement. Specifically, the relation between $m^\prime$ and $a$ under the condition $b$ is deterministic and therefore reversible if (and only if) \cite{Hof12}
\begin{equation}
\label{eq:determinism}
\sum_a P(m|a,b) P(a|m^\prime,b) = \delta_{m,m^\prime}.
\end{equation}
This relation explains why the complex conditional probabilities (or action phase probabilities) obtained from weak measurements are universal expressions of determinism. They are in fact universal representations of the fundamental relations between physical properties that applies to all states and measurements. 

\section{What ``superposition'' really means}

Much confusion has been caused by the description of quantum states as ``superpositions'' of other states. It is therefore important to understand the physics that is actually expressed by Hilbert state vectors and their components. In fact, the components of Hilbert space vectors should be understood as expressions of unitary transformations generated by the basis $\{\mid m \rangle \}$ used to expand the initial state $\mid a \rangle$. Specifically, the dynamics of transformations is expressed by changes in the complex phases of the components, so that the change in the probability of $b$ is given by
\begin{eqnarray}
P(U(b)|a) &=& |\langle b \mid \hat{U} \mid a \rangle|^2
\nonumber \\
&=& \left|\sum_m \langle b \mid m \rangle \langle m \mid a \rangle e^{-i S(m)/\hbar}\right|^2,
\end{eqnarray}
where $S(m)$ is the action of the transformation generated by $m$. If $a$ and $b$ overlap, it is possible to replace the Hilbert space vectors with action phase probabilities \cite{Hof11},
\begin{equation}
P(U(b)|a) = P(b|a) \left|\sum_m P(m|a,b) e^{-i S(m)/\hbar}\right|^2
\end{equation}
Primarily, the components of state vectors therefore express the action of transformations, and not the probabilities of measurement outcomes. This is actually part of the definition of the phases in Hilbert space, since the seemingly arbitrary choice of phase zero for each component in a $d$-dimensional Hilbert space is actually defined by the physics of the reference state
\begin{equation}
\mid r \rangle = \frac{1}{\sqrt{d}} \sum_m \mid m \rangle.
\end{equation}
With respect to this reference state, any physical property $a$ can be described by its action phase probabilities $P(m|a,r)$, where 
\begin{equation}
\label{eq:vector}
P(m|a,r) = \frac{\langle m \mid a \rangle}{\sum_{m^\prime} \langle m^\prime \mid a \rangle}.
\end{equation}
As shown in \cite{Hof14}, it is possible to derive the complete Hilbert space formalism from the action phase probabilities that describe the deterministic relations between any three physical properties. In particular, the relation (\ref{eq:vector}) above shows that the Hilbert space vector of a state is really an expression of the transformations between the respective state and a reference state $r$ generated by the basis property $m$. Thus the components of the vector represent the dynamics of the state, and not its ``realities''. 

Quantum ergodicity is the insight that randomness in quantum mechanics has a dynamical origin. The uncertainty principle of quantum state preparation is actually a consequence of the specific form of causality observed in the limit of precisely defined actions, and the state vector is a representation of these causality relations with respect to an implicit reference state $r$.

\section{Quantum environments and cosmological concerns}

The essential insight of quantum ergodicity is that quantum mechanics applies equally to all objects, whether they are microscopic or macroscopic. It is a mistake to think that the description of an environment or a measurement apparatus can be ``classical''. Therefore, the laws of physics that determine state preparation must be the same as the laws of physics that determine the unitary evolution between preparation and measurement. The approach succeeds because the closed system formalism of quantum mechanics can be formulated in terms of universal relations between physical properties. When applied to state preparation, these relations identify the microscopic quantum fluctuations in the environment as the origin of randomness in the quantum state. 

In the conventional formalism, it is suggested that interactions entangle the microscopic properties of the environment with the properties of the quantum system. State preparation then requires an additional ``measurement'' of the environment. What I am arguing here is that this conventional viewpoint is mistaken, because our actual relation with the physical environment does not work in this way. Empirical reality is controllable only at the most macroscopic level, and we need to accept that this lack of control is a fundamental part of the world we live in. We should realize that simple everyday manipulations already involve the relations of quantum physics. For example, the precise alignment of a beam splitter corresponds to a well-defined quantum coherence of its angular momentum, and the associated fluctuations supply the angular momentum that depolarizes the circular polarization that passes through the beam splitter. 

This principle of ergodic randomness is even more valid in large systems than it is in small ones. Most large systems interact with their environment through a specific selection of properties that are particularly easy to observe and thus qualify as ``macroscopic'' properties. However, the laws of physics require that all properties are dynamically connected with conjugated properties. Therefore, the macroscopic properties of large objects are dynamically connected to microscopic properties that are extremely difficult to observe because they interact only very weakly and are rapidly randomized by the interactions that make the macroscopic properties so easy to see. What appears as a ``classical'' reality is therefore always connected to a reservoir of quantum fluctuations that will be relevant in any interactions between the large system and a microscopic system. 

Microscopic properties only obtain reality through macroscopic effects. In the conventional formalism, both state preparation and measurement are represented by summaries of the ergodic averaging that occurs in the interactions between the quantum system and the macroscopic devices that enable us to control the system. The limits that quantum theory imposes on this control are actual limits of physical reality, as described by the deterministic relations between physical properties.  

It may be important to emphasize that the principle of quantum ergodicity is not just a pragmatic form of instrumentalism, where the output of the device is real while the original object is not. The formulation of quantum physics in terms of action phase probabilities removes boundary between the system and its environment completely, so there is no division into ``object'' and ``device''. It is possible to apply this approach directly to quantum cosmology. When investigating the universe, the objective properties we observe are quite macroscopic. We can take any property of interest and construct a dynamically connected generator of transformations. It should be easy to confirm that this property has no observable effects at cosmological scales. In principle, these cosmological quantum fluctuations can be traced back all the way to the big bang, where they were already included as a specific correlation between the physical properties at the beginning of the universe. 

The point is that most of the microscopic degrees of freedom present at the big bang never had any macroscopic consequences that would be visible in the universe as it appears to us today. At the quantum scale of $\hbar$, the universe is a poorly controlled system indeed, with much more randomness attributed to thermal fluctuations than to quantum fluctuations - both of which are of course equally ergodic. 

\section{Conclusions}

The randomness of quantum systems has its identifiable origin in the dynamics of quantum state preparation. The reason why quantum mechanical randomness appears to be so different from classical randomness is that it is conditioned by the {\it dynamics} of quantum systems. In the preparation of a pure state, the necessary interaction randomizes the dynamics of transformations that keep the intended property constant. The probability distributions $P(b|a)$ given by the squared inner product $|\langle b \mid a \rangle|^2$ can all be explained as a result of the ergodic randomization of $b$ along a trajectory of $a$. 

The difference between quantum ergodic dynamics and classical dynamics is described by complex probabilities that replace the classical deterministic relations between physical properties with conditional probabilities $P(m|a,b)$, where the complex phases of the probabilities represent the action of an optimized transformation from $a$ to $b$ along $m$ \cite{Hof11}. The appearance of quantum coherence in the Hilbert space formalism can thus be explained in terms of universal laws of causality that describe the response of quantum systems to external forces. 

``Classical'' reality emerges by approximation, when the relations between physical properties are only determined with a precision that is much lower than $\hbar$. This approximation applies to most of our experience, since large objects are characterized by only a few macroscopic properties and a much larger number of microscopic properties that provide an inexhaustible supply quantum ergodic fluctuations. 

Statistical models of quantum mechanics result in paradoxes because the fundamental relations between physical properties are described by action phase probabilities. These complex-valued probabilities describe statistical correlations that cannot be explained by assigning positive probabilities to joint realities of the physical properties involved. Instead, the dynamical relation between the physical properties explains why it is not possible to control all physical properties of a system at the same time: control is fundamentally limited by the laws of causality that govern all interactions of the system with its macroscopic environment. 

Ultimately, a proper formulation of quantum physics should address these issues by explaining the laws of causality first. The statistics of states can then be derived from an analysis of the process by which states are prepared.



\begin{thebibliography}{10}


\bibitem{Wis07}
H. M. Wiseman, New J. Phys. {\bf 9}, 165 (2007).

\bibitem{Hal14}
M. J. W. Hall, D.-A. Deckert, and H. M. Wiseman,
Phys. Rev. X {\bf 4}, 041013 (2014). 

\bibitem{Bar12}
S. D. Bartlett, T. Rudolph, and R. W. Spekkens,
Phys. Rev. A {\bf 86}, 012103 (2012).


\bibitem{Hof11}
H.F. Hofmann, New J. Phys. {\bf 13}, 103009 (2011).

\bibitem{Hof12}
H.F. Hofmann, New J. Phys. {\bf 14}, 043031 (2012).

\bibitem{Hof14}
H. F. Hofmann, Phys. Rev. A {\bf 89}, 042115 (2014).


\bibitem{BasDal}
For the Stern-Gerlach experiment, a very good explanation of the origin of state preparation noise is found in the textbook ``Quantum Mechanics'' by J.-L. Basvedart and J. Dalibard, Springer 2002, section 8.6.

\bibitem{Bol1898}
Part 2 of ``Vorlesungen \"uber Gastheorie'' by L. Boltzmann, published in 1898 by J. A. Barth, Leipzig.

\bibitem{Aha88}
Y. Aharonov, D.Z. Albert, and L. Vaidman, Phys. Rev. Lett. {\bf 60}, 1351 (1988).


\bibitem{Lun11}
J.S. Lundeen, B. Sutherland, A. Patel, C. Stewart, and C. Bamber, Nature {\bf 474}, 188 (2011).


\end{thebibliography}
\end{document}